\documentclass[pre,onecolumn,showpacs,preprintnumbers,amsmath,amssymb]{revtex4}
\usepackage{graphicx,graphics,bm}

\begin{document}

\title{Optimum and efficient sampling for variational quantum Monte Carlo}

\author{J. R. Trail}\email{jrtrail@jaist.ac.jp}
\author{Ryo Maezono}
\affiliation{School of Information Science, Japan Advanced Institute of Science 
and Technology, Ishikawa, Japan}

\date{June, 2010}

\begin{abstract}
Quantum mechanics for many-body systems may be reduced to the evaluation of integrals in 3N dimensions using Monte-Carlo, providing the Quantum Monte Carlo \emph{ab initio} methods.

Here we limit ourselves to expectation values for trial wavefunctions, that is to Variational quantum Monte Carlo.
Almost all previous implementations employ samples distributed as the physical probability density of the trial wavefunction, and assume the Central Limit Theorem to be valid.
In this paper we provide an analysis of random error in estimation and optimisation that leads naturally to new sampling strategies with improved computational and statistical properties.
A rigorous lower limit to the random error is derived, and an efficient sampling strategy presented that significantly increases computational efficiency.
In addition the infinite variance heavy tailed random errors of optimum parameters in conventional methods are replaced with a Normal random error, strengthening the theoretical basis of optimisation.

The method is applied to a number of first-row systems, and compared with previously published results.
\end{abstract}



\maketitle

\section{Introduction}

The accurate computational solution of the many-body Schr\"odinger equation for fermionic systems is an important and only partially solved problem in modern chemistry.
Monte-Carlo (MC) integration provides one of the most powerful methods available to address this problem, and has been implemented in many different ways\cite{foulkes01}.
All methods take a reformulation of the problem in terms of multi-dimensional integrals, such as a variational principle or propagators, and employ MC methods for the evaluation of the integrals involved.
The strength of MC integration is that it provides superior scaling of accuracy with the number of integrand evaluations.
For example, for $r$ evaluations of a smooth integrand the MC error scales as $\sim 1/r^{1/2}$ independent of the dimensionality of the system ($D$), whereas for order-$p$ polynomial interpolation on an evenly sampled grid the error scales as $\sim(1/r)^{p/D}$ \cite{traub98}.

In this paper we limit ourselves to the Variational quantum Monte Carlo (VMC) method.
Expectation values for trial wavefunctions are estimated directly, with a trial wavefunction constructed to explicitly include correlation and for which the dimensionality of integrals cannot be reduced analytically.
Typically a trial wavefunction is expressed in terms of a large number of unknown parameters, and a variational principle invoked to provide the `best' possible trial wavefunction.
The optimised quantity is estimated using MC.
Higher accuracy Quantum Monte Carlo (QMC) methods are available, but VMC itself provides a high level of accuracy, is often the starting point of other QMC methods, and other QMC methods include MC integration in a similar manner.

For MC integration a key choice is the distribution of random samples used to construct the statistical estimate.
Such a choice is not unique, and different choices result in different distributions of the random error inherent in the estimate of the integral.
For a MC estimate to be \emph{useful} the underlying distribution from which it is drawn must be known both in form and scale via the limit theorems of probability theory.
This is not always possible.
Within VMC the characterisation of random errors is further complicated by the physical quantities being quotients of estimated integrals.

In light of this the main body of this paper is concerned with characterising the statistical properties of estimates that arise from different choice of sampling strategy, and falls into three main sections.
We start by considering estimates of expectation values only, with no optimisation of parameters, and for general sampling.
A large class of distributions of random error are shown to arise, the most desirable being the Normal distribution that occurs where the Central Limit Theorem (CLT) is valid in its bivariate form.
Conditions are provided for when this occurs, and estimates are provided for the mean and variance of this Normal distribution.
Requiring a Normal random error still allows a wide range of sampling distribution, and alternatives to the usual method used in the VMC literature (referred to here as `standard sampling') are discussed.
An optimum sampling distribution is derived that provides a lower bound to the random error possible for a given sample size, and a benchmark for assessing the effectiveness of other sampling strategies.
An efficient sampling distribution is constructed with the aim of significantly reducing the computational resources required to achieve a given random error.
For both optimum and efficient sampling errors are Normal.

Next we address the issue of trial wavefunction optimisation.
A number of methods exist for achieving an iterative improvement of an initial wavefunction within VMC, which may be interpreted as seeking the minimum (or zero) of a multivariate function that is random in the sense that it is drawn from an ensemble of possible random functions.
The methods used to find the optimum of a sample function are of secondary importance, and we focus on the distribution of the position of the minimum of the random function, itself a random variable.
This viewpoint naturally divides the optimisation process into taking a sample random function and numerically optimising to some level of accuracy (a `cycle' of optimisation), then taking a new sample random function and optimising again (repeated `cycles' of optimisation).
We discuss the distributions of random error in the sample optimum trial wavefunction, how much information limit theorems provide, and the performance of different sampling strategies.

Finally, the efficient sampling strategy is used for the optimisation and estimation of the electronic energies of first row isolated atoms and molecules, with the systems chosen for comparison with previously published \emph{ab initio} results and accurate approximations for the exact energies.
Our aim is to demonstrate the quality of results that can be obtained using new sampling strategies in VMC, complicated trial wavefunctions, and modest computational resources.

Sans-serif is used for random variables such as $\mathsf{X}$, with a sample value denoted as $X$.
The expectation of a random variable is denoted $\mathbb{E}[\mathsf{X}]$, and a random estimates of some quantity $y$ (usually an integral or a quotient of two integrals) is denoted $\widehat{y}$, and is a sample value of some random variable $\mathsf{Y}$.
The integration variable of a probability density function (PDF) is usually left implicit, but where it is explicit we use $x$, $y$ or $u$ for scalar, and $\mathbf{R}$ for vector arguments.
Atomic units are used throughout, unless otherwise indicated.

\section{Generalised sampling}
\label{sec:gen}

We begin with the expectation value of the total energy, $E_{tot}$, of an $N$ electron system (a generalisation to different operators and particles in straightforward).
For a given Hamiltonian and trial wavefunction, $\hat{H}$ and $\psi$, this takes the form
\begin{equation}
E_{tot}=\int \psi^2 E_L[\psi] d\mathbf{R} / 
        \int \psi^2           d\mathbf{R} ,
\end{equation}
where the local energy is defined as $E_L[\psi] = \psi^{-1} \hat{H} \psi$, and $\mathbf{R}$ is the $3N$ dimensional vector composed of the positions of all electrons.

A general sampling approach to VMC consists of defining a random position vector $\mathsf{R}$ characterised by some distribution and drawing $r$ independent samples of this random variable.
Writing the distribution as 
\begin{equation}
P_g= \lambda \psi^2 / w,
\end{equation}
is convenient in what follows, with $w$ a general (positive) weight function defined in the $3N$ dimensional space of the integrals.
The normalisation prefactor, $\lambda$, is not required to generate samples using the Metropolis algorithm\cite{foulkes01}, and is included in order to clarify that expressions for estimates are invariate with respect to it.
Note that it is unrelated to the physical normalisation of the trial wavefunction unless $w=1$.

Estimating both integrals using MC provides an estimates of the total energy as
\begin{equation}
\widehat{E}_{tot} = \frac{ \sum w E_L }{ \sum w },
\label{eq:1}
\end{equation}
but for this estimate to be useful its distribution must be known.
Since this quantity is the quotient of two sample means the CLT is not directly applicable.

The sample weights in this equation may be renormalised  such that their sum is equal to the number of samples by introducing a new weight, $w'=r w/\sum w$, so reducing the number of random variables by a factor of two and reducing Eq.~(\ref{eq:1}) to the sample mean
\begin{equation}
\widehat{E}_{tot}= \frac{1}{r}\sum{w'E_L}.
\end{equation}
Unfortunately a price is payed for this apparent simplification of the problem.
In order for the CLT to be valid (at least in its simplest form) it is necessary for the samples to be independent and identically distributed.
For the reformulation given above neither of these conditions are satisfied since each sample value of $w'$ is composed of the all of the original samples, $\{w\}$.
Although a sample mean and sample standard error exist the CLT does not relate these to the mean and variance of a Normal distribution, or ensure that the distribution is Normal.

A more successful approach is to consider the estimate explicitly as a quotient of two sums, hence maintaining the independence and identical distribution of the random variables sampled in each sum, and including the parametric relationship between each sample $wE_L$ and $w$.
We begin by obtaining the co-distribution of the numerator and denominator of Eq.~(\ref{eq:1}), so include all correlation between them.
The bivariate CLT informs us that in the large sample size limit $(r^{-1}\sum w E_L , r^{-1}\sum w)$ is a sample drawn from a bivariate normal distribution\cite{stroock93} with PDF
\begin{equation}
P_r(x_2,x_1)=\frac{1}{2\pi} \frac{r^{1/2}}{| C |^{1/2}} e^{-1/2q^2}
\end{equation}
characterised by a mean, $(\mu_2,\mu_1)$,
\begin{eqnarray}
(\mu_2,\mu_1)= \left(  \mathbb{E}\left[  wE_L \right] ,  \mathbb{E}\left[  w    \right] \right),
\end{eqnarray}
a covariance matrix, $C$, with elements 
\begin{eqnarray}
c_{22}=\text{Var}\left[ wE_L \right]          ,&
c_{12}=c_{21}=\text{Cov}\left[ wE_L,w \right] ,&
c_{11}=\text{Var}\left[ w    \right]          ,  
\end{eqnarray}
and
\begin{equation}
q^2= r
\begin{pmatrix} x_1 - \mu_1  \\
                x_2 - \mu_2  \end{pmatrix}^{T}
C^{-1}
\begin{pmatrix} x_1 - \mu_1  \\
                x_2 - \mu_2  \end{pmatrix} .
\end{equation}
Although the elements of each sample $(wE_L,w)$ are causally related to each other, the only aspect of this that survives in the large sample size limit is the partial linear correlation characterised by $c_{12}$.

From this bivariate Normal distribution the distribution of the quotient of sums can be expressed using the standard formulae\cite{curtiss41}
\begin{equation}
P_r(u)= -\int_{-\infty}^{0} x_1 P_r( x_2=ux_1, x_1)  dx_1
        +\int_{0}^{+\infty} x_1 P_r( x_2=ux_1, x_1)  dx_1,
\end{equation}
which, in the large sample size limit, reduces to 
\begin{eqnarray}
 P_r\left( u \right)&=&
\frac{r^{1/2}}{\sqrt{2\pi}}
\left|
\frac{
\left( c_{11}\mu_2-c_{12}\mu_1 \right)u + \left( c_{22}\mu_1-c_{12}\mu_2 \right)
}{
\left( c_{11}u^2-2c_{12}u+c_{22} \right)^{3/2}
}
\right|
\nonumber \\ & & \times
\exp{ \left[
-\frac{r}{2}
\frac{ \left( \mu_2-\mu_1u \right )^2           }
     { \left( c_{11}u^2-2c_{12}u+c_{22} \right) }
\right] }
\end{eqnarray}
for general $\mu_1$.
Since the weights are necessarily positive, $\mu_1 > 0$ and the large sample size limit further reduces to 
\begin{equation}
P_r(u)=\frac{1}{\sqrt{2\pi}} \frac{1}{\sigma_E} \exp{ \left[ -\frac{ (u - E_{tot})^2 }{2 \sigma_E^2} \right] },
\end{equation}
a Normal distribution with mean
\begin{equation}
E_{tot} = \frac{ \int \psi^2 E_L               d\mathbf{R} }
               { \int \psi^2                   d\mathbf{R} }.
\label{eq:2}
\end{equation}
and variance
\begin{eqnarray}
\sigma_E^2  &=& \frac{1}{r}\frac{1}{\mu_1^2} \left[ 
                 c_{22} 
                - 2 \left(\frac{\mu_2}{\mu_1}\right)c_{12} 
                + \left(\frac{\mu_2}{\mu_1}\right)^2c_{11}             \right] \nonumber \\
            &=& \frac{1}{r}
                \frac{   \int \psi^2/w d\mathbf{R}   \int w \psi^2 (E_L - E_{tot})^2 d\mathbf{R}   }
                     {   \left[ \int \psi^2 d\mathbf{R} \right]^2  }.
\label{eq:3}
\end{eqnarray}

Replacing the means by their unbiased estimates provides the total energy estimate
\begin{equation}
\widehat{E}_{tot} = \frac{ \sum w E_L }{ \sum w },
\label{eq:4}
\end{equation}
as a sample drawn from a Normal distribution with a mean equal to the exact total energy expectation value.
The variance is defined by Eq.~(\ref{eq:3}), and introducing standard unbiased estimates for the means and covariance matrix elements gives the standard error estimate as
\begin{equation}
\widehat{\sigma_E^2} = \frac{r}{r-1}
\frac{ \sum w^2 \left( E_L - \widehat{E}_{tot} \right)^2 }
     {          \left( \sum w                  \right)^2 }.
\label{eq:5}
\end{equation}
The usual interpretation of the accuracy of the estimates in terms of confidence limits and standard errors then follows directly, provided we use these two formulae.

We finish with some comments on the properties and limitations of these estimates.
The estimates cannot be derived using the univariate CLT, and the standard error is not a sample variance divided by a sample size.
For $P_g$ to be a valid PDF it is necessary for $\psi^2/w$ to be positive and normalisable.
For the bivariate CLT to be valid the integral definitions of the mean vector and covariance matrix must exist, which cannot be shown numerically - the situation is analogous to the assumptions of smoothness that are required to provide error limits for numerical integration on a grid.
Perhaps most importantly, for an exact trial wavefunction $\widehat{\sigma_E^2} = 0$ for any number of samples and for any choice of sampling distribution, hence generalised sampling possesses the `zero variance' property in that there is no statistical error in the total energy estimate of an eigenfunction of the Hamiltonian.

The above equations inform us that estimated expectation values are Normal provided that the covariance exists.
In the rest of this paper we repeatedly seek estimates for which this is the case, hence here we summarise why this is desirable and what replaces the Normal distribution when the covariance is not defined.

The \emph{generalised} CLT\cite{nolan10} states that, in the large $r$ limit, the average of independent and identically distributed random variables drawn from a distribution $P$ possesses either a Normal distribution or a Stable distribution. (Throughout this paper we refer to a Normal distribution as distinct from a Stable distribution.
This is for brevity, and Stable distributions are often defined to include the Normal distribution as a special case.)
Which of these arises depends on the properties of $P$.
For tails of $P$ decaying faster than $|x|^{-3}$ the strongly localised Normal distribution occurs which is characterised by a mean and variance that can be estimated.
For tails of $P$ decaying slower than $|x|^{-3}$ a weakly localised Stable distributions occurs which is characterised by 4 parameters which cannot all be estimated.
For the Normal distribution the PDF is Gaussian and the width of the distribution scales as $r^{-1/2}$, whereas for Stable distributions the PDF asymptotically approaches a power law and its width scales slower than $r^{-1/2}$, is constant, or increases with $r$ depending on the tails of $P$.

\begin{figure}[t]
\includegraphics[scale=1.00]{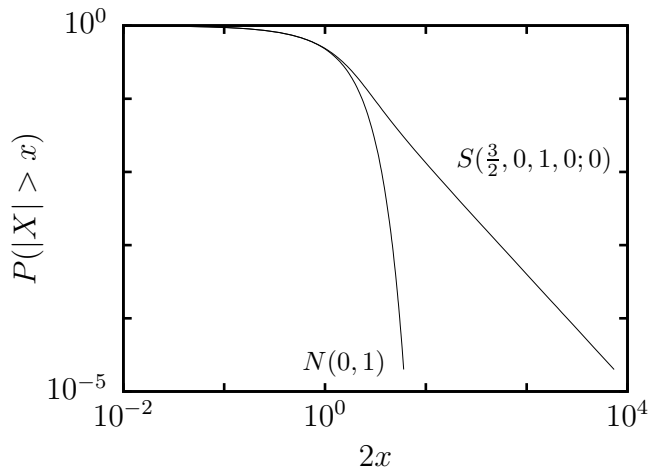}
\caption{ \label{fig0} 
Probability that a sample falls outside of a centralised interval of size $2x$ for $N(0,1)$, a Normal distribution, and for an example Stable distribution, $S(\frac{1}{2},0,1,0 ; 0)$\cite{nolan10} that is symmetric with $x^{-5/2}$ power law tails and mean $0$ (the mean exists for this Stable distribution).
This form of Stable distribution is typical of the distributions that occur for an invalid CLT in standard sampling QMC.
}
\end{figure}

The Stable case is undesirable for two reasons.
Estimates are not available for the parameters of the distribution, hence characterising the error via confidence intervals is not straightforward.
Furthermore, even if parameters could be estimated, the presence of power law tails causes the size of the confidence interval to increase with the level of confidence considerably faster than the CLT case.
An example of this is shown in Figure~\ref{fig0}.
The probability that a sample falls outside of a centred region is shown as a function of the size of the region, for a Normal distribution and for a Stable distribution.
The Normal distribution is chosen to have a mean and variance of zero and unity.
The Stable distribution shown is symmetric, has $|x|^{-5/2}$ tails, has a mean that exists and is equal to zero, and has a `scale' parameter (the analogue of the variance for the Normal case) chosen to be unity.
There is no relationship between the variance and `scale' of the two distributions, and they are chosen to be equal in this figure for convenience only.
Such a probability asymptotically approaches $e^{-\frac{1}{4}x^2}/x$ for the Normal case, with outliers remaining persistent for the Stable distribution due to the $|x|^{-3/2}$ asymptote.
Where an estimate is composed of a quotient of sample averages and covariances are not defined a similar situation arises;  a bivariate generalised CLT is required, a bivariate Stable distribution results\cite{nolan10}, power law tails arise and estimates of error bars are unavailable.

The notion of using a sampling distribution such that total energy estimates require weighted averages is inherently part of the re-weighted sampling of optimisation, Diffusion Monte Carlo (DMC), and other QMC methods, but is usually analysed by incorrectly applying the univariate CLT to numerically renormalised samples.
Explicit generalised sampling in VMC has been used by a few authors\cite{dewing02,coldwell79,alexander91,sorella08}, and Coldwell first presented a formula for the variance estimate that differs from Eq.~(\ref{eq:5}) only in that it does not include the prefactor that corrects for bias in the estimated covariance matrix.
Similarly, Assaraf \emph{et al.}\cite{assaraf07} employ the same expression in the context of DMC, again without the correction prefactor.

The new results provided here are a proof that Eq.~(\ref{eq:4}) is an estimate that is Normally distributed provided that the covariance matrix exists, and that if this is so the mean of the distribution is the total energy and Eq.~(\ref{eq:5}) estimates the variance with the appropriate bias correction.
Note that requiring the covariance to exist refers to the integral definition, not to sample estimates since the later are always finite.

Next we consider particular choices of sampling distribution, $P_g$.

\subsection{Standard sampling}

Choosing $w=1$ over all space corresponds to the usual implementation of VMC, sampling the co-ordinate space as 
\begin{equation}
P_{s}=\lambda \psi^2
\end{equation}
which gives the estimate
\begin{equation}
\widehat{E}_{tot} = \frac{1}{r} \sum E_L
\end{equation}
with standard error
\begin{equation}
\widehat{\sigma_E^2} = \frac{1}{r(r-1)} \sum \left( E_L - \widehat{E}_{tot} \right)^2.
\end{equation}
Perhaps the greatest advantage of this choice is the simple algebraic form of the estimates, and that they may be derived using the univariate CLT only.
However, the previous consideration of generalised sampling suggests that lower random errors may be possible for other sampling distributions, and that similar random errors may occur for other sampling distributions.

The standard sampling choice also has some further deficiencies, as discussed in a previous paper\cite{trail08a}, due to the necessary existence of a $3N-1$ dimensional nodal hyper-surface where $\psi$ is zero.
Such zeroes are usually accompanied by singularities in the sampled local energy on these surfaces (this is not necessarily the case, indeed is not the case for an exact wavefunction, but is usually unavoidable), and the analogous `local' quantities averaged to estimate expectation values for other operators.
Such singularities are purely an artifact of zeroes in $P_s$, and cause the distribution of estimates for many physical quantities to be a Stable distribution with power law tails.
For standard sampling the total energy estimate is somewhat unusual in that CLT is valid\cite{trail08a,trail08b}.
For most other quantities, such as the derivatives used for forces and in wavefunction optimisation, this is not so.

\subsection{Optimum sampling}

For a given trial wavefunction and Hamiltonian, the random error depends on sample size and the sample distribution characterised by the weight function $w$.
The random error for a given system and number of samples must be greater than or equal to zero, hence a weight function must exist that supplies the smallest possible random error in an estimate.
So, the general standard error defined by Eq.~(\ref{eq:3}) should possess minimum with respect to variations in the weight.
Seeking $\delta \sigma_E^2\left[w\right] / \delta w=0$, gives 
\begin{equation}
 w^2=\frac{1}{ (E_L-E_{tot})^2 }
     \frac{ \int \psi^2 w (E_L-E_{tot})^2 d\mathbf{R} }
          { \int \psi^2/w d\mathbf{R} },
\end{equation}
whose solution (excluding $w=0$) gives the optimum sampling distribution,
\begin{eqnarray}
P_{opt}=\lambda \psi^2 | E_L - E_{tot} |,
\label{eq:6}
\end{eqnarray}
which is consistent with the optimal sampling distribution for estimating the energy differences between two systems provided by Ceperley \emph{et al.}\cite{ceperley02}.

Equation~(\ref{eq:6}) provides a lower bound to the achievable accuracy in that $\sigma_E \geq \sigma_{opt}$ for all sampling strategies, with 
\begin{eqnarray}
\sigma_{opt} &=& \frac{1}{ r^{\frac{1}{2}} } \int \psi^2 |E_L - E_{tot}| d{\bf R},
\end{eqnarray}
a scaled mean-absolute deviation of the local energy.
This is related to the more familiar standard error of standard sampling,
\begin{eqnarray}
\sigma_{s} &=& \frac{1}{ r^{\frac{1}{2}} } \left[ \int \psi^2 (E_L - E_{tot})^2 d{\bf R} \right]^{1/2},
\end{eqnarray}
by $\sigma_{opt} \leq \sigma_{s}$ (from Jensen's inequality).
Estimates are readily available for both of these integrals.

This lower limit is analytically correct, but does not provide a satisfactory distribution for the implementation of optimum sampling.
For the weight defined in Eq.~(\ref{eq:6}) the covariance matrix is undefined unless $E_L \ne E_{tot}$ for any $\mathbf{R}$, a special condition that we cannot reasonably expect to be generally satisfied.
Consequently the bivariate CLT will not be valid for most systems. \cite{foot1}

Even worse, evaluating the PDF requires the exact $E_{tot}$, which is generally not available as it is the quantity whose value is sought by the MC calculation.
A natural approximation is to replace the exact $E_{tot}$ with some previous non-exact estimate. 
Unfortunately this worsens the statistics considerably, resulting in an `estimate' that does not statistically converge to a constant with increasing sample size.

An improved definition of optimum sampling is required.
We take a self consistent approach in order to allow $E_{tot}$ to be unknown, and postpone the large sample size limit as late as possible in the derivation in order to ensure that the covariance matrix exists.
Fiellers's theorem\cite{luxburg04} provides a convenient reformulation of the distribution of a quotient of bivariate Normal random variables in terms of the confidence intervals, $[l_l,l_u)$.
These intervals are defined by the concise expression 
\begin{equation}
l_{l,u} = \frac{
\left( r \mu_1.\mu_2-q^2.c_{12} \right) \pm
   \sqrt{\left( r\mu_1.\mu_2-q^2.c_{12} \right)^2 -
         \left( r\mu_1^2              -q^2.c_{11} \right)
         \left( r\mu_2^2              -q^2.c_{22} \right)
        } }{    r\mu_1^2              -q^2.c_{11} }
\end{equation}
for confidence $\textrm{erf}\left( q/\sqrt{2} \right)$, for example $q=1$ provides the $68.3\%$ confidence interval.
As long as $\mu_1>0$ such a description is entirely equivalent to the Normal distribution specified by Eqs.~(\ref{eq:2},\ref{eq:3}) since in the large $r$ limit it defines the interval $E_{tot}\pm q.\sigma_E$ for confidence $\textrm{erf}\left( q/\sqrt{2} \right)$.

Minimising the size of the confidence interval $2.\sigma_F=l_u-l_l$ using $\delta \sigma_F^2 /\delta w=0$ provides
\begin{equation}
0=2\sigma_F^2
  \left(1-\frac{1}{r} \frac{1}{\mu_1^2} c_{11} \right)
  \frac{\delta  }{\delta w} \left( c_{11} \right)
 +\frac{1}{r} \frac{1}{\mu_1^2} \frac{\delta  }{\delta w} \left( c_{12}^2-c_{11}c_{22} \right)
 +\frac{\delta  }{\delta w} \left( c_{22} -2 E_{tot} c_{12} + E_{tot}^2 c_{11} \right).
\end{equation}
Solving the large $r$ limit of this equation reproduces Eq.~(\ref{eq:6}), but instead we approximate this equation by introducing an \emph{a priori} total energy estimate and error that are constant with respect to $r$, that is replace $(E_{tot},\sigma_F)$ with a $(E_0,\epsilon)$.
The large $r$ limit then gives
\begin{equation}
0= 2 \epsilon^2 \frac{\delta  }{\delta w} \left( c_{11} \right)
 +\frac{\delta  }{\delta w} \left( c_{22} -2 E_{0} c_{12} + E_0^2 c_{11} \right),
\end{equation}
and solving for $w$ gives 
\begin{eqnarray}
P_{opt}=\lambda \psi^2 \left[ (E_L - E_0)^2 + 2 \epsilon^2 \right]^{1/2}.
\label{eq:7}
\end{eqnarray}
This distribution is optimum in the sense that it is the solution of the best available approximation to the equation that defines optimum sampling.

In principle the estimates could be calculated iteratively by supplying $(E_{tot},\sigma_F)$ from one calculation as $(E_0,\epsilon)$ for the next, and proceeding to convergence.
Since $(E_0,\epsilon)$ would not converge to the exact estimate, the covariance matrix remains defined and the bivariate CLT is valid.
In practical application we go no further than the first iteration, using relatively inaccurate total energies such as those provided by an optimisation process.

A poor value of $(E_0,\epsilon)$ does not bias the total energy estimate, which converges to the true value as long as $\epsilon \neq 0$.
Similarly, accurate values for $E_0$ are not necessary as long as this is reflected in the accompanying error, $\epsilon$, in that for large $\epsilon$ this sampling strategy becomes equivalent to standard sampling.
However, it should be borne in mind that the random error can be greater than that for standard sampling if $\epsilon$ underestimates the accuracy of $E_0$.

As an aside, we note that it is possible to construct a sampling distribution that provides a better value of the total energy than this optimum sampling strategy, but such sampling strategies are not MC estimates.
For example, a $P_g$ for which all the probability is located on the hyper-surface $E_L=E_{tot}$ is such a distribution, but neither the mean or the co-variance matrix are defined and the `estimate' is entirely made up of bias.
Hence the algorithm simply returns the value of $E_{tot}$ supplied at the start.
Although the optimum strategy described above does requires some estimate of the total energy (and indeed its error) this does not skew the distribution or introduce a systematic bias.

 Sampling as Eq.~(\ref{eq:7}) is referred to as optimum sampling in what follows, as it provides the best available approximation to the lowest possible random error for a given sample size.

\subsection{Efficient sampling}

Optimum sampling provides a useful lower limit to the statistical error achievable for a given system, and in practise it is possible to get very close to this limit using relatively inaccurate estimates of $(E_0,\epsilon)$.
However, optimum sampling will not necessarily be efficient as no account has been taken of the computational cost of generating each of the samples, and we seek a further sampling strategy for which the statistical error available for a given computational cost is low.

For VMC, the algorithm usually used to draw samples from a distribution is the generation of a Markov chain via the Metropolis algorithm.
This is well documented in the literature\cite{foulkes01} and can provide an ordered list of sample vectors drawn from a list of correlated random vectors each distributed as $P_g$, and for which the degree of correlation between pairs in the list falls exponentially as they become more separate.
Disregarding enough of these samples results in a new list for which correlation is negligible, providing (effectively) independent and identically distributed random vectors.
This is the set $\{\mathbf{R}\}_r$ used to construct MC estimates.

Consequently, the computational cost of generating a VMC estimate is composed of evaluating $P_g$ at $mr$ accept/reject steps and evaluating $E_L$ at $r$ of these steps.
The total computational cost is $T = mT_{P_g} + T_{E_L}$ where $T_{P_g}$ is the computational cost of evaluating $P_g$, and $T_{E_L}$ is the computational cost of evaluating the local energy, suggesting that the available accuracy may be improved by reducing $T_{P_g}$ and increasing the total number of samples, $r$ \cite{foot2}.
Our goal is to seek an `efficient' sampling strategy for which the PDF is computationally cheap to evaluate, for which the bivariate CLT is valid, and for which the standard error for a given number of samples is low.
As a pragmatic measure of success we use the statistical accuracy achievable for a fixed computational budget.

A simple, but not generally successful, approach is to approximate standard sampling by choosing a single slater determinant $\Phi_{SD} \approx \psi$ (such as a Hartree Fock (HF) Slater determinant) and to sample using
\begin{eqnarray}
P_{SD}=\lambda \Phi_{SD}^2.
\end{eqnarray}
For the associated estimates to be useful the bivariate CLT must be valid, and appendix \ref{sec:appa} describes the analysis used to ascertain whether this is so, essentially a more general form of that described in Ref.~\cite{trail08a}.
To summarise, both of $(wE_L,w)$ possess a singularity at the nodal surface of $\Phi_{SD}$ that cause the covariance of $(wE_L,w)$ to be undefined.
Consequently, each sample mean is itself a sample drawn from a Stable distribution with a mean, with $x^{-5/2}$ power law tails, and with no variance.
Although an estimate constructed using Eq.~(\ref{eq:4}) is computationally cheap and does converge to the total energy expectation value in the large sample size limit, no information is available about the distribution of errors.
An important exception occurs if the nodal surfaces of $\psi$ and $\Phi_{SD}$ coincide, for which the bivariate CLT is valid, the total energy estimate is Normal, and the standard error can be estimated.

In light of this we seek a sample distribution that does not possess a nodal surface, yet shares many of the properties of the trial wavefunction and is arithmetically simple in comparison with both standard sampling and optimum sampling.
This is not enough to uniquely define a sampling distribution, and we propose the form
\begin{eqnarray}
P_{eff} = \lambda \left[ |\Phi_1(\mathbf{R})|^2+|\Phi_2(\mathbf{R})|^2 \right],
\label{eq:8}
\end{eqnarray}
where $\Phi_1$ and $\Phi_2$ are Slater determinants (or Configuration State Functions) with different nodal surfaces.
A source of such determinants might be the dominant components of a multi-determinant wavefunction, or a ground and excited state supplied by HF.

The distribution of Eq.~(\ref{eq:8}) has the desirable qualities that it is computationally cheap to evaluate relative to many of the trial wavefunctions used in QMC, is not zero on any $3N-1$ dimensional nodal surface, is zero on the $3N-3$ dimensional coalescence hyper-planes, and it reproduces exponential tails for electrons in bound systems.
A physical interpretation of this expression would be somewhat tenuous, and not particularly useful, and it should be considered solely as a compromise between analytic properties and ease of calculation.
The analysis of appendix \ref{sec:appa} informs us that no singularities are present in either of $(wE_L,w)$, power law tails do not arise, and both the mean vector and covariance matrix exists.
Consequently, this distribution provides a total energy estimate that is Normally distributed and whose variance can be estimated.
Sampling using Eq.~(\ref{eq:8}) is referred to as efficient sampling in what follows.

To compare the performance of efficient, standard, optimum and SD sampling total energy estimates were calculated for an isolated all-electron Carbon atom in the $^3P$ non-relativistic ground state.
A version of the CASINO\cite{casino10} package was used, with modifications introduced to enable generalised sampling.
Of the variety of trial wavefunctions available for use in QMC we employ the form 
\begin{equation}
\psi= e^{J(\bf R)} \sum_{n=1}^{N_{\rm CSF}} \alpha_{n} \Phi_{n}( {\bf X} )
\label{eq:wf}
\end{equation}
that is made up of a multi-determinant expansion, Jastrow prefactor, and backflow transformation, hence allows variation of the nodal surface.
The Jastrow factor, $J(\bf R)$ includes electron-electron, electron-ion, and electron-electron-ion terms (see Ref.~\cite{drummond04}).
Backflow is included via the non-linear transformation $\bf R \rightarrow \bf X(\bf R)$ made up of electron-electron, electron-ion, and electron-electron-ion terms (see Ref.~\cite{rios06}).
Configuration State Functions (CSFs) for the multi-determinant expansion, $\Phi_{n}$, are obtained from the multi-configurational Hartree-Fock (MCHF) atomic-structure package \textsc{ATSP2K}\cite{fischer07}, and are made up of numerical orbitals and a reduced active space that includes CSFs in order of increasing weight.
To summarise, the Jastrow factor is characterised by $167$ parameters, backflow is characterised by $161$ parameters, and the $550$ determinant expansion of $52$ CSFs is characterised by $51$ parameters.
This trial wavefunction is of essentially the same form as Ref.~\cite{brown07}, but with the number of parameters systematically increased.
Samples were taken from the metropolis random walk every $m=64$ steps, a value chosen as the lowest power of $2$ for which correlation was statistically undetectable.

Values for wavefunction parameters were obtained using optimisation as described in the next section.
We ask the reader to bear with this somewhat inverted order, since a valid sampling strategy for estimates is a necessary prerequisite for optimisation even though in application optimisation must be performed first.

\begin{figure}[t]
\includegraphics[scale=1.00]{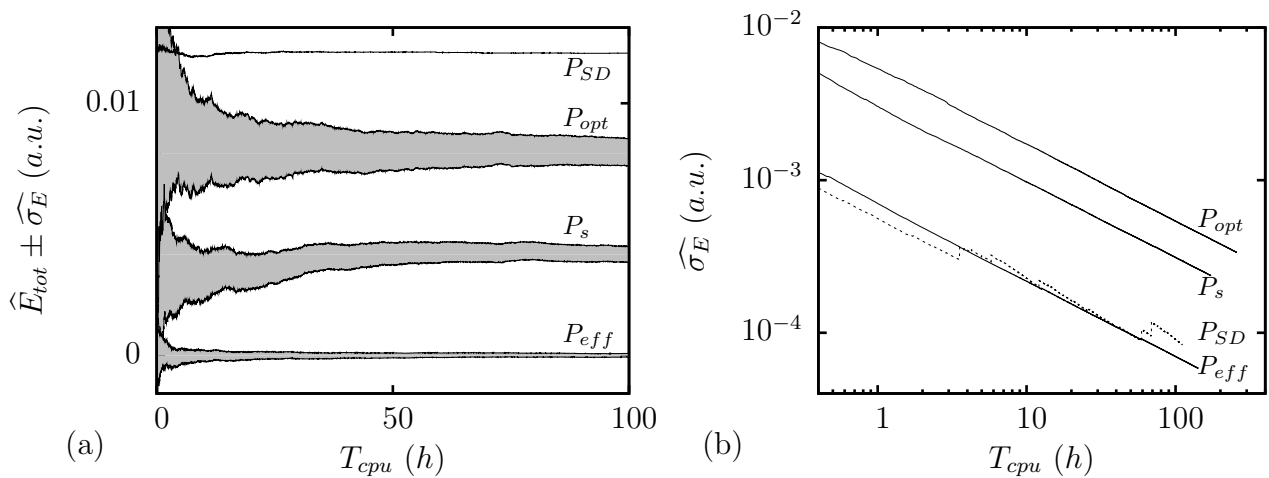}
\caption{ \label{fig1} 
Evolution of total energy estimates with computational time.
Figure (a) shows the evolution of confidence intervals for efficient, standard and optimum sampling. No confidence interval is available for SD sampling, and only the estimated value is shown. An arbitrary constant offset has been added for each sampling strategy to aid clarity.
Figure (b) shows the evolution of standard errors for efficient, standard and optimum sampling.
For SD sampling the quantity shown arises from evaluating the standard error estimate when it is incorrect to do so, and is unrelated to confidence intervals.
}
\end{figure}

For a fixed sample size there is a modest variation in the standard error (Eq.~(\ref{eq:5})).
When compared with standard sampling, efficient sampling increases the error by $29\%$, optimum sampling reduces the error by $37\%$, and SD sampling provides no standard error.
However, computational cost does vary significantly.
Figure~\ref{fig1}a shows the evolution of total energy estimates with computational time.
The standard error for a given computational cost is lowest for efficient sampling, and greatest for optimum sampling, with no confidence interval available for SD sampling (only the estimate is shown).
The SD sampling estimate does appear to approach a constant value, but this is likely to be illusory since estimates for different sample sizes are strongly correlated.

Figure~\ref{fig1}b shows the evolution of the estimated error with computational time for all four sampling strategies, with errors obtained from Eq.~(\ref{eq:5}).
Efficient, standard and optimum sampling show a consistent $\propto T_{cpu}^{-1/2}$ evolution with a large variation in the prefactor between different sampling strategies.
For SD sampling the result of an incorrect application of Eq.~(\ref{eq:5}) shows no consistent power law behaviour with sporadic discontinuous jumps introduced by outliers, and although confidence intervals do exist for the underlying distribution they are not related to the quantity shown in the figure.
Sample sizes and estimates obtained by allocating a fixed computational resource to each sampling strategy are given in Table~\ref{tab:1}.
Compared to standard sampling, efficient sampling increases the number of samples by $\times 34$, reducing the error by $\times 1/5$, whereas optimum sampling decreases the number of samples by $\times 1/7$, increasing the error by $\times 2$.

\begin{table}[b]
\begin{tabular}{lrl}                              \hline \hline
\multicolumn{1}{c}{Type} & \multicolumn{1}{c}{$r$} & \multicolumn{1}{c}{$\widehat{E}_{tot}$}  \\ \hline
$P_{opt}$  &    430096 & -37.8435(2)           \\
$P_{s}$    &   3193192 & -37.8436(1)           \\
$P_{eff}$  & 106829024 & -37.84361(2)          \\
$P_{SD}$   & 136971968 & -37.843607            \\ \hline \hline
\end{tabular}
\caption{ \label{tab:1} Total energy estimates for fixed computational time and optimum, standard, efficient and SD sampling. No standard error is available for SD sampling and the number of significant figures is unrelated to random error.
}
\end{table}

The performance improvement of efficient sampling when compared to standard sampling varies between systems, and some limits are available.
For $r$ samples distributed as $P_{eff}$, and assuming that the cost of evaluating the PDF is negligible when compared with evaluating the trial wavefunction, the speedup with respect to standard sampling is the ratio
\begin{equation}
\frac{T_{s}}{T_{eff}} \approx (m-1) \frac{T_{\psi^2}}{T_{E_L}} + 1.
\end{equation}
Since $T_{\psi^2}/T_{E_L} < 1$ the speedup is limited to less than $m$-fold, with few electron systems closest to this limit.
The opposite limit occurs as the number of electrons in the system is increased, where $T_{\psi^2}/T_{E_L} \rightarrow 0$ or no speedup occurs.
For all cases speedup increases with $m$, with none occurring for $m=1$.
How close we are to each of these limits depends strongly on the relative computational cost of wavefunction evaluation and local energy evaluation (which includes wavefunction evaluation) so must be considered on a case by case basis.

For example, for the Li, C, and Ne$_2$ calculations performed later in this paper standard sampling requires $\times 22$, $\times 20$, and $\times 5$ more computational resource to achieve the same accuracy as efficient sampling.

\section{Optimisation}
\label{sec:opt}

In the previous section the trial wavefunction, $\psi$, was taken as given and the random error in total energy estimates considered.
In this section $\psi$ is taken to be variable, and we address the role of randomness in the process of searching for optimum parameters using MC estimates.
This can be viewed as extending random MC from estimates of single valued quantities (such as the total energy) to estimates of smooth functions (such as total energy as a function of wavefunction parameters), and seeking the minimum or zero of such a random function.

It has long been appreciated that using independent VMC estimates for different parameter values results in a discontinuous surface, hence the powerful methods available for continuous surfaces are not directly applicable.
This is usually avoided by using `correlated sampling'\cite{foulkes01}, where a set of sample positions, $\{\mathbf{R}\}_r$, is generated from a VMC calculation using initial parameters, and a MC estimate derived or proposed that provides a continuous sample function drawn from an underlying ensemble of functions and whose minimum may be found using standard numerical methods.
Cycles of optimisation are repeated with fresh sample functions, some convergence criteria is applied, and the final set of parameters are supplied to a final VMC estimate of expectation values.

This introduces random error into the final estimate in two distinct ways.
The final VMC estimate for a given set of parameters has a random error as described in the previous section.
However, the parameters used are also samples drawn from a random distribution (they are the positions of the zero of a random function) and introduce a random error that is additive since optimisation and estimation are independent of each other.

For the specific example of total energy, the estimates are sample values of 
\begin{equation}
\mathsf{E}_{opt} = E_{tot} + \mathsf{e}_{vmc} + \mathsf{e}_{opt},
\end{equation}
where $E_{tot}$ is the exact optimum total energy allowed by the variational freedom of the trial wavefunction,
$\mathsf{e}_{vmc}$ is the random error in a VMC total energy estimate performed with the optimised parameters, and
$\mathsf{e}_{opt}$ is the random error due to the optimisation being performed on a random function.
If no MC estimation was required in the optimisation process then $\mathsf{e}_{opt}=0$, but $\mathsf{e}_{vmc}$ would remain non-zero.
Similarly, if the final total energy evaluation was exact $\mathsf{e}_{vmc}=0$, but MC optimisation would still give a non-zero $\mathsf{e}_{opt}$.

The random variable $\mathsf{e}_{vmc}$ was discussed in the previous section, and is well understood in that it is possible to ensure that it is Normal and to estimate a standard error for it.
However, $\mathsf{e}_{opt}$ cannot be dealt with in the same way since it involves the random variation in the shape of the sample function.
It would be useful obtain some information about the distribution of $\mathsf{e}_{opt}$, how it varies with the number of samples and parameters, and what limit theorems are applicable to it.
In what follows we present a preliminary analysis that provides some of this information.

It is worthwhile to note that optimisation to numerical convergence within a cycle takes a large number of function evaluations, and most of this computational effort is wasted since the optimised function itself has a random error.
This can be easily avoided by relaxing convergence criteria, for example limiting the number of iterations allowed within each cycle can significantly improve computational efficiency.

In what follows we denote the number of samples used for estimation and optimisation as $r_{vmc}$ and $r_{opt}$, with $r_{vmc} \ge r_{opt}$ due to the sample electronic positions used in optimisation being supplied by a VMC estimate calculation.

\subsection{Optimisation with a sample function}

A variety of functions are estimated for use in VMC optimisation.
Here we provide a brief description their form, their distribution, and how they generalise to non-standard sampling.

The family of methods loosely referred to as `variance minimisation' seek a minimum of a function, $g(\mathbf{a})$, of the general form 
\begin{eqnarray}
g(\mathbf{a}) &=& \int f. (E_L(\mathbf{a})-E)^2 d\mathbf{R} \left/
                  \int f. d\mathbf{R} \right.                    \nonumber \\
E             &=& \int f.  E_L(\mathbf{a}) d\mathbf{R} \left/
                  \int f. d\mathbf{R} \right.
\end{eqnarray}
characterised by $f$, a non-negative function of co-ordinate, $\mathbf{R}$, and by the trial wavefunction characterised by parameters $\mathbf{a}$.

For general sampling, the MC estimate used to define the optimisation surface takes the form \cite{foot3}
\begin{eqnarray}
\widehat{g}(\mathbf{a}) &=& 
\frac{ \sum w(\mathbf{a}) (E_L(\mathbf{a})-\widehat{E})^2 }
     { \sum w(\mathbf{a}) }                                      \nonumber \\
\widehat{E}         &=& \frac{\sum w(\mathbf{a}) E_L(\mathbf{a}) }{ \sum w(\mathbf{a}) } ,
\label{eq:9}
\end{eqnarray}
where the weight and distribution are related by $P_g=\lambda f/w$ with $P_g$ independent of $\mathbf{a}$.
The estimated function $\widehat{g}(\mathbf{a})$ has a lower bound of zero, and obeys a zero variance principle in that both the estimate and its gradient are exactly zero for an exact trial wavefunction.
These two properties allow standard numerical methods to be stable and successful.
However, the estimate of the random surface is not necessarily Normally distributed.

Standard sampling implementations of such methods correspond to choosing $P_g=\lambda \psi^2(\mathbf{a}_0)$ where $\mathbf{a}_0$ are the initial parameters of a cycle of optimisation.
For this case the analysis of appendix \ref{sec:appa} informs us that for most choices of $f$ the bivariate CLT fails, with the distribution of $\widehat{g}(\mathbf{a})$ possessing power law tails, no variance, and in some cases no mean.
The closely related `mean absolute deviation' optimisation functions can be obtained by replacing the squared deviation in Eq.~(\ref{eq:9}) with an absolute deviation \cite{alexander91}, and infinite variance distributions arise for the same reasons.

For either case a Normal distribution of errors occurs only for `artificial weighting', where $f$ is modified such that it approaches zero at the nodal surface sufficiently fast to remove singularities in the sampled quantity.
However, this approach does not seem entirely satisfactory, since the failure of the CLT is due to standard sampling, not to properties of the integral whose value is estimated.
Furthermore, more advanced QMC methods specifically require trial wavefunctions with accurate nodal surfaces, hence artificially suppressing the contribution of this region to the optimised quantity is not desirable.

Minimisation of the total energy expectation value for a trial wavefunction is perhaps the most desirable choice as it is involves a physical variational principle.
Consequently we focus on total energy optimisation by generalising the `linear optimisation method'\cite{nightingale01,toulouse07} to generalised sampling, and provide a statistical analysis of the method.

The total energy estimate for samples distributed as $P_g = \lambda \psi^2(\mathbf{a}) / w(\mathbf{a})$ (with $P_g$ independent of $\mathbf{a}$) is given by
\begin{eqnarray}
\widehat{E}_{tot}(\mathbf{a}) &=& \frac{ \sum w(\mathbf{a}) E_L(\mathbf{a}) }{ \sum w(\mathbf{a}) } ,
\end{eqnarray}
a sample surface drawn from a distribution of random surfaces.
Optimisation on this surface is usually unsuccessful since it is not bounded from below, and that although the energy estimate itself is `zero-variance' its gradient is not.

Hermiticity of the Hamiltonian operator naturally provides a variety of estimates for the total energy gradient, including a zero-variance estimate.
The cost of choosing the zero-variance gradient estimate is that its gradient (the Hessian) is not a symmetric matrix, hence the estimated gradient is not itself a derivative of a sample surface.

For notational simplicity we introduce a new parameter vector, $\mathbf{b}=(b_0,a_1,\ldots)^{T}$, that includes a normalisation factor via $\psi(\mathbf{b})=b_0 \psi(\mathbf{a})$ to give the generalised sampling zero-variance gradient estimate as
\begin{eqnarray}
\widehat{E}^{(1)}_i[\psi(\mathbf{b})]  &=& 2\left. 
    \frac{ \sum w \psi^{-1} \psi_i \left[ E_L[\psi] - \widehat{E}_{tot}[\psi] \right] }
         { \sum w } \right|_{\mathbf{b}}
\label{eq:10}
\end{eqnarray}
where gradients in parameters space are denoted as $E^{(1)}_i = ( \nabla E(\mathbf{b})    )_i$ and $\psi_i =  ( \nabla \psi(\mathbf{b}) )_i$, and parameters have been made implicit.

Linear optimisation proceeds by seeking the zero of Eq.~(\ref{eq:10}) using successive 1st order expansions of the trial wavefunction.
Expansion about parameters $\mathbf{b}^n$ provides this as
\begin{equation}
\tilde{\psi}^{n+1} (\mathbf{b}^n + \mathbf{\Delta b}^{n+1} ) 
=  \mathbf{\Delta b}^{n+1}. \nabla \psi^n (\mathbf{b}^n).
\end{equation}
Replacing $\psi$ with $\tilde{\psi}^{n+1}$ in the condition $\widehat{E}^{(1)}_i[\psi(\mathbf{b})]=0$ gives
\begin{equation}
\widehat{\mathbf{H}}[\mathbf{b}^n] \mathbf{\Delta b}^{n+1} = 
   \epsilon^n \widehat{\mathbf{S}}[\mathbf{b}^n] \mathbf{\Delta b}^{n+1}
\end{equation}
an eigenvalue equation with matrix elements given by the MC estimates
\begin{eqnarray}
\widehat{H}_{ij} = \left. \sum w \psi_0^{-2} \psi_i \psi_j E_L[\psi_j] \right|_{\mathbf{b}^n} & , &
\widehat{S}_{ij} = \left. \sum w \psi_0^{-2} \psi_i \psi_j \right|_{\mathbf{b}^n} .
\label{eq:11}
\end{eqnarray}
Solving for the lowest eigenvalue provides updated parameter values,
\begin{eqnarray}
\mathbf{b}^{n+1} &=& \frac{ 1 } { b^{n}_0 + \Delta b^{n+1}_0 } ( \mathbf{b}^{n} + \mathbf{\Delta b}^{n+1} ).
\end{eqnarray}
If this process is iterated to self-consistency, then
\begin{eqnarray}
\widehat{E}^{(1)}_i[\psi(\mathbf{b}^{n+1})] & = & 0 \nonumber \\
\mathbf{\Delta b}^{n+1} &=& (1,0,\ldots)^{T}        \nonumber \\
\epsilon_{n+1} &=& \left. \frac{ \sum w E_L[\psi] }{ \sum w } \right|_{\mathbf{b}_n},
\end{eqnarray}
that is the parameters converge to those for which the estimated gradient is zero.

To ensure that such a method is robust we use the semi-orthogonalisation, rescaling, and level-shift modifications as developed by Toulouse \emph{et al.}\cite{toulouse07}, which easily transfer to generalised sampling, and normalise the diagonal of $\widehat{S}_{ij}$ to unity to provide an energy scale for level-shift parameters.
Note that these modifications do not alter the self-consistent solution, but are necessary to stabilise the method.

The convergence behaviour of this approach requires some attention.
For the special case where the variational freedom of the trial wavefunction includes the exact eigenstate, then it is clear that self-consistency occurs for a zero gradient estimate and a minimum energy estimate at the same parameter values.
However, the two will not generally coincide since the gradient estimate is not usually the gradient of the total energy estimate.
Furthermore, the gradient estimate is not the derivative of a sample surface so there is no reason to expect a point to exist where the gradient is zero.
So, self-consistency need not occur.
Limiting the optimisation to a few (usually one) iteration within each cycle and monitoring convergence across cycles provide an effective control for this indefiniteness.
In what follows we do not take this into account explicitly, and assume that its effect can be entirely accounted for by allowing for correlation between cycles and a modification of $\mathsf{e}_{opt}$ that becomes negligible close to statistical convergence.

The distribution of errors of the estimated gradient function (whose zero is sought), and of the estimated matrix elements can be assessed as described in appendix \ref{sec:appa}.
For standard sampling, Stable distributions arise for estimates with a mean, with $x^{-\frac{5}{2}}$ power law tails, and with no variance.
For the gradient estimate this occurs even for a stationary nodal surface.
The total energy estimate performs better, in that it is Normal for a stationary nodal surface and for the initial parameters of a cycle, but not otherwise.
Similarly, most of the matrix elements are samples drawn from a Stable distribution with $x^{-\frac{5}{2}}$ tails.

Employing efficient or optimum sampling of section \ref{sec:gen} removes the nodal surface in the sampling distribution, hence estimates of the total energy, total energy gradient, and matrix elements are all Normal. \cite{foot4}

\subsection{Random errors in optimisation}

Given the distribution of the gradient estimate we may characterise the optimisation random error, $\mathsf{e}_{opt}$.
In principle the route towards this is straightforward.
For both variance and linear optimisation of the total energy we seek a zero value of a smooth yet random gradient, that is solutions of the random equation
\begin{equation}
\text{\sffamily\bfseries g}^{(1)}( \text{\sffamily\bfseries a}_{opt}  ) = 0.
\label{eq:12}
\end{equation}
Samples of this random vector function are defined by $\widehat{\mathbf{g}}^{(1)} = \nabla \widehat{g}( \mathbf{a} )$ (see Eq.~(\ref{eq:9})) for variance minimisation, and by $( \widehat{\mathbf{g}}^{(1)} )_i = \widehat{E}^{(1)}_i$ (see Eq.~(\ref{eq:10})) for linear optimisation.
Solving the random equation defines the `random optimum' parameters, $\text{\sffamily\bfseries a}_{opt}$, and solving the associated estimate of the gradient equation provides sample values of the `random optimum' parameters.
Introducing these random parameters into the total energy then defines the random optimisation error as
\begin{equation}
\mathsf{e}_{opt} = E_{tot}\left[ \text{\sffamily\bfseries a}_{opt} \right] 
                 - E_{tot}\left[ \mathbf{a}_{0} \right]
\end{equation}
where $\mathbf{a}_{0}$ is the exact minimum of the total energy with respect to parameter variation.
It is immediately apparent that $\mathsf{e}_{opt} \geq 0$ and $\mathbb{E}[\mathsf{e}_{opt}] \geq 0$.
It is also clear that the random error in optimum parameters is related to the random error in the derivative of the minimised function, not to the random error in the function itself.

Deriving the distribution of $\mathsf{e}_{opt}$ is not simple, with the primary difficulty due to inversion of the random vector function, and here we offer a perturbative solution.
We begin by expanding the random vector function about the true minimum, $\mathbf{a}_{0}$,
\begin{equation}
\text{\sffamily\bfseries g}^{(1)}(\mathbf{a})= 
\text{\sffamily\bfseries g}^{(1)}(\mathbf{a}_0) + 
\text{\sffamily\bfseries g}^{(2)}(\mathbf{a}_0) (\mathbf{a}-\mathbf{a}_{0}) + \ldots
\label{eq:13}
\end{equation}
where each coefficient in the expansion is a random variable.
Truncating to 1st order and solving for $\text{\sffamily\bfseries g}^{(1)}=0$ gives the random deviation from the exact optimum as
\begin{equation}
\Delta \text{\sffamily\bfseries a} =
\text{\sffamily\bfseries a}_{opt} - \mathbf{a}_{0} = -
  \left.
  \left[ \text{\sffamily\bfseries g}^{(2)} \right]^{-1}.
  \text{\sffamily\bfseries g}^{(1)}
  \right|_{\mathbf{a}_{0}}.
\end{equation}
To simplify this expression further we rotate and rescale parameters such that $\mathbb{E} \left[ \text{\sffamily\bfseries g}^{(2)}(\mathbf{a}_{0}) \right]=\mathbf{I}$, and expand the matrix/vector product to 1st order in the deviation of each random variable from its mean value to give
\begin{equation}
\Delta \text{\sffamily\bfseries a} =
  \left.
 -\text{\sffamily\bfseries g}^{(1)}
 +\left[  \text{\sffamily\bfseries g}^{(2)} - \mathbf{I}  \right]
  \mathbb{E} \left[ \text{\sffamily\bfseries g}^{(1)} \right]
  \right|_{\mathbf{a}_{0}},
\label{eq:14}
\end{equation}
a linear combination of random variables.

Expanding the exact total energy to 2nd order about the true minimum and introducing these `random optimum' parameters gives
\begin{eqnarray}
\mathsf{e}_{opt} = 
\frac{1}{2} \Delta \text{\sffamily\bfseries a}^T . \mathbf{I} . 
            \Delta \text{\sffamily\bfseries a},
\end{eqnarray}
a sum of the squares of $N_p$ correlated random variables.
Provided that both $\text{\sffamily\bfseries g}^{(1)}$ and $\text{\sffamily\bfseries g}^{(2)}$ possess a mean and a variance then it follows that $\mathsf{e}_{opt}$ possesses a mean and a variance given by
\begin{eqnarray}
\mathbb{E}\left[ \mathsf{e}_{opt} \right] &=& 
    \frac{1}{2} \sum_i \mathbb{E}\left[ \Delta\mathsf{a}_i^2 \right]   \nonumber \\
\text{Var}\left[ \mathsf{e}_{opt} \right] &=&
    \frac{1}{4}\sum_{ij} \left( 
    \mathbb{E}\left[\Delta\mathsf{a}_i^2\Delta\mathsf{a}_j^2\right]-
    \mathbb{E}\left[\Delta\mathsf{a}_i^2\right]\mathbb{E}\left[\Delta\mathsf{a}_j^2\right] 
    \right)
\label{eq:15}
\end{eqnarray}
where $\Delta\mathsf{a}_i=(\Delta \text{\sffamily\bfseries a})_i$.
These equations allow us to deduce some the properties of the distribution of $\mathsf{e}_{opt}$ for different sampling distributions.

For both efficient and optimum sampling, $\text{\sffamily\bfseries g}^{(1)}(\mathbf{a})$ is Normal, hence
$\text{\sffamily\bfseries g}^{(1)}(\mathbf{a}_0)$, $\text{\sffamily\bfseries g}^{(2)}(\mathbf{a}_0)$, and $\Delta \text{\sffamily\bfseries a}$ are multivariate Normal.
Writing the covariance matrix elements of this distribution as $C_{ij}/r_{opt}$ (they arise from the bivariate CLT as discussed in the previous section) and limiting ourselves to total energy minimisation (where the mean of the random deviation is zero) then gives the co-moments\cite{lax06}
\begin{eqnarray}
\mathbb{E}\left[\Delta\mathsf{a}_i\right]                       &=& 0                             \nonumber \\
\mathbb{E}\left[\Delta\mathsf{a}_i^2\right]                     &=& C_{ij}/r_{opt}                \nonumber \\
\mathbb{E}\left[\Delta\mathsf{a}_i^2\Delta\mathsf{a}_j^2\right] &=& (C_{ii}C_{jj}+2C_{ij}^2)/r_{opt}^2 ,
\end{eqnarray}
which may be introduced into Eq.~(\ref{eq:15}) to give the first two moments of $\mathsf{e}_{opt}$ as
\begin{eqnarray}
\mathbb{E}\left[ \mathsf{e}_{opt} \right] &= (2r_{opt})^{-1}   \sum_i    C_{ii}    \nonumber \\
\text{Var}\left[ \mathsf{e}_{opt} \right] &= (2r_{opt}^2)^{-1} \sum_{ij} C_{ij}^2  .
\end{eqnarray}
Provided that correlation between each $\Delta\mathsf{a}_i$ is weak enough the distribution of $\mathsf{e}_{opt}$ will be a generalised $\chi^2$ distribution and approach Normal in the large $N_p$ limit, but we do not assume this to be the case.

This analysis informs us that, provided the mean and variance exist, the optimisation error can be systematically decreased by employing more samples.
However, increasing $N_p$ causes two conflicting effects.
The number of co-variances in each sum increases, contributing a linear increase in the optimisation error.
In competition to this all covariances are reduced by the extra variational freedom in the trial wavefunction.
The second of these effects will not necessarily dominate over the first, hence added variational freedom will not necessarily improve VMC results.

For standard sampling Normality is absent from the first stage of the above analysis, since $\text{\sffamily\bfseries g}^{(1)}(\mathbf{a})$ is not Normally distributed and possesses no variance.
So, the `random optimum' parameters are not Normally distributed, Eqs.~(\ref{eq:15}) are undefined, and the distribution of $\mathsf{e}_{opt}$ possesses neither a mean or a variance.
Although we cannot conclude that it possesses a Stable Distribution, we can conclude that the distribution from which it is drawn possesses power law tails that decay as $x^{-2}$ or slower, and that it approaches a $\delta(x)$ function with increasing $r_{opt}$. 

A number of points should be made to clarify this analysis.
Exact quantities used above are not available for an actual calculation, but this does not influence the conclusion since we only require these quantities to exist.
Similarly, the definition of parameters such that the exact hessian is the identity matrix only simplifies notation.
A more serious issue is that the above analysis is not a proof due to the truncation of series expansions to 1st and 2nd order.
However, this perturbative approach is often successful for Normal random variables, hence the analysis can be considered as strongly suggestive.

To summarise, the random error due to optimisation is greater than zero, may increase with increasing number of parameters, and is unrelated to the random error in the total energy estimate for a single set of parameters.
For efficient or optimum sampling, the mean and standard deviation of this optimisation error both scale as $r_{opt}^{-1}$.
For standard sampling no such result arises and the optimisation error is drawn from an unknown distribution with no mean or variance.

Since optimisation with efficient (or optimum) sampling provides more control over random errors this suggests that it will perform better than standard sampling.
The analysis also suggest that (for efficient or optimum sampling) the statistical quality of parameters can be improved by performing cycles of optimisation until convergence is achieved, and averaging the parameters provided by a $n$ further cycles, so reducing the mean and standard deviation of $\mathsf{e}_{opt}$ by a factor of $1/n$.
In a sense this is equivalent to optimising with $nr$ samples and not averaging, but has the advantage of allowing faster convergence that can be monitored.

\begin{figure}[t]
\includegraphics[scale=1.00]{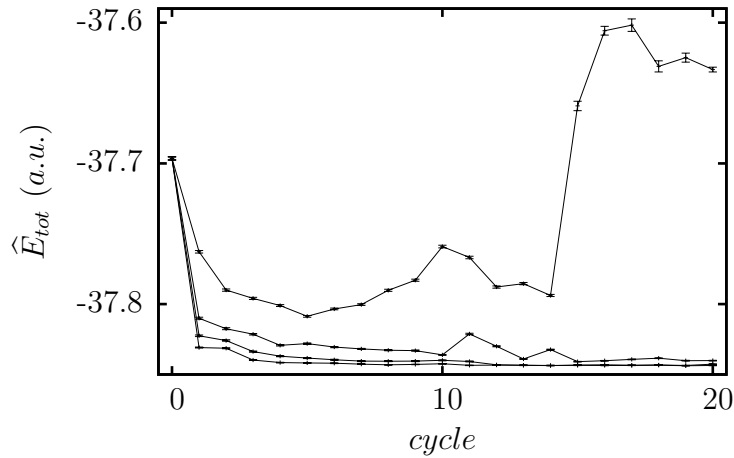}
\caption{ \label{fig2}
Evolution of total energy estimates with optimisation for all-electron C, and using efficient sampling.
Four optimisation processes are shown using 384, 1480, 14800, and 80000 samples in order of decreasing $\widehat{E}_{tot}$.
All total energy estimates are generated using 2200000 samples.
}
\end{figure}

We consider optimisation of the isolated all-electron Carbon atom described in section \ref{sec:gen} using efficient sampling.
Figure \ref{fig2} shows the evolution of estimates during optimisation using efficient sampling.
Twenty cycles of optimisation are shown, with $r_{opt}$ varying between $384$ and $80000$ (there are 377 free parameters), and with total energy estimates evaluated from $r_{vmc}=2200000$ samples.
As $r_{opt}$ is increased, the convergence rapidly becomes stable.
Estimates arising from from $r_{opt}=14800$ and $r_{opt}=80000$ are statistically indistinguishable, and for both the final 9 estimates possess $97.3\%$ confidence intervals that overlap with the confidence intervals of the lowest energy estimate.
Essentially the $\mathsf{e}_{opt}$ becomes undetectable in the presence of $\mathsf{e}_{vmc}$ for $r_{vmc} \sim 150r_{opt}$, an equal allocation of computational resource to VMC estimation and optimisation.

We expect $\mathsf{e}_{opt}$ to be worse for standard sampling, in that power law tails and outliers occur.
Exploratory calculations show this to be so for small sample sizes, for no robust stabilisation, or for including several iterations in each cycle.
But, for modest sample sizes, including robust stabilisation, and employing only one iteration per cycle, the optimisation error remains undetectable at the lower statistical resolution of standard sampling.
It seems that the primary advantages offered by efficient sampling is the improved accuracy for estimates and the theoretical justification for averaging converged sets of parameters.

\begin{figure}[t]
\includegraphics[scale=1.00]{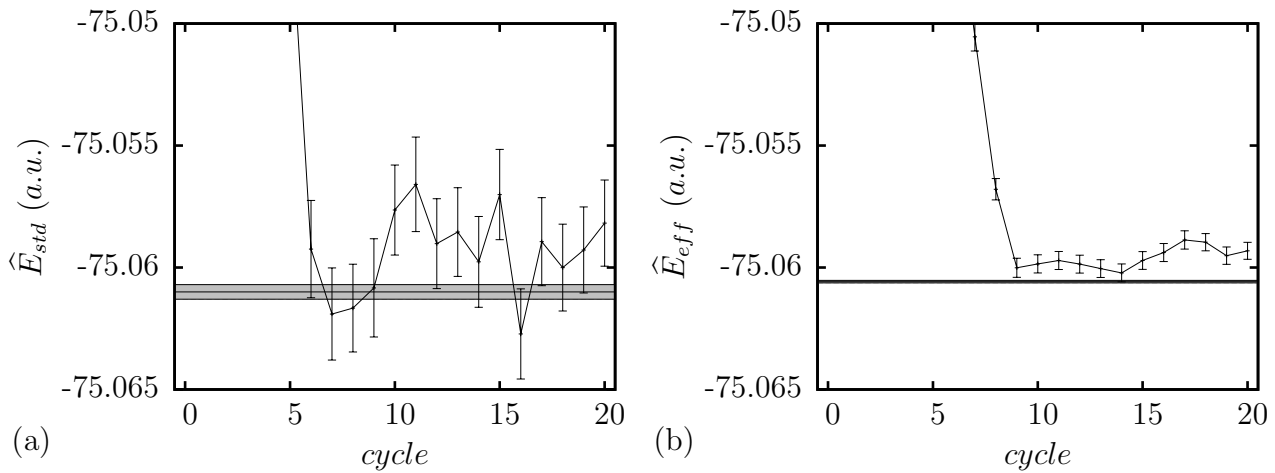}
\caption{ \label{fig3} 
Evolution of total energy estimates with optimisation for O, for both standard and efficient sampling.
Figure (a) shows the evolution of estimates using standard sampling, and figure (b) shows the evolution of estimates using efficient sampling, with the total computational cost of each the same.
The grey regions are the confidence interval for the final total energy estimate constructed using an average of converged parameters and equal computational cost.
}
\end{figure}

Figure \ref{fig3} show the results of standard and efficient optimisation for an isolated O atom, with the same computational resource used for each.
The final VMC total energy estimate was constructed using $50\%$ of the resource, $25\%$ was used to generate samples and total energy estimates for cycles, and $25\%$ was used for the optimisation itself.
This corresponds to samples proportioned between final estimate, monitor estimate and optimisation as $220:5:1$ for standard sampling and $8946:186:1$ for efficient sampling.

Efficient sampling performs significantly better than standard sampling for all the total energy estimates, primarily by reducing the random error by $\times1/7$.
For efficient sampling the final estimate is consistently below the less accurate estimates arising in optimisation, but such an improvement is not clearly discernible for standard sampling.
We take this combination of efficient sampling, equipartition of computational cost, and averaging of statistically indistinguishable sets of parameters and apply it to generate the results of the next section.

\section{Results and discussion}
\label{sec:res}

The approach was applied to a group of first row systems, the isolated atoms, homonuclear diatomic molecules, diatomic hydrides, and the diatomic molecules LiF, CN, CO, and NO.
This collection of systems was chosen to allow comparison of total energy estimates with recently published \emph{ab initio} total energies and accurate approximate total energies in the literature.

For any VMC calculation the achievable statistical and systematic error is unavoidably a compromise between finite computational resources and the incompleteness of available wavefunction parametrisation.
To make our compromise we chose to allocate the same computational resource to each system, divided between optimisation and estimation as in the previous section \cite{foot5}.
Trial wavefunctions of the form described in section \ref{sec:opt}, Eq.~(\ref{eq:wf}), were constructed with expansion orders chosen to include as much variational freedom as possible whilst obtaining a usefully low statistical error.

For both Jastrow and backflow the number of parameters were set so as to employ polynomials 2-orders less than the onset of numerical error in polynomial evaluation.
Multi-determinant expansions were constructed as sums of uncoupled spatial and spin symmetry eigenstates, that is as Configuration State Functions (CSFs), with optimised coefficients.
For isolated atoms the CSF expansion was obtained using numerical orbitals from \textsc{ATSP2K}\cite{fischer07} as for C in section \ref{sec:gen}.
For the diatomic molecules no MCHF package appears to be available, hence 
ground state and some low energy excited state numerical orbitals were generated using the Hartree-Fock package \textsc{2dhf}\cite{laaksonen86}, and CSFs constructed from these using single and double excitations of valence electrons.
A somewhat arbitrary upper limit to the number of CSFs was used, aiming for 500 and 50 determinants for isolated atoms and molecules, but with a large amount of variation in the later due to the finite number of excited bound states.

These trial wavefunctions offer a large degree of variational freedom, with between 246 and 455 optimisable parameters present (for Ne$_2$ and BeH respectively).
The most obvious deficiencies are that orbital relaxation is not included, and that for the diatomic molecules the chosen multi-determinant expansion is not self consistent.
Table~\ref{tab:2} summarises the properties of each trial wavefunction.
Optimisation and estimation of these trial wavefunction proceeded as in section \ref{sec:opt}, with $20$ cycles of optimisation followed by averaging of converged parameters for use in the final total energy estimate calculation.
Overall optimisation showed similar stability to that shown previously for C and O, providing between 1 and 10 converged sets of parameters for averaging.

\begin{table}[t]
\begin{tabular}{lrccrrr}    \hline \hline
       & $N_e$ & \ d\ (a.u.) \ & Term & $N_{det}$  & $N_{CSF}$  & $N_{param}$   \\ \hline
Li     &  3  & -        & $^2 S         $ & 502 & 86 & 326 \\
Be     &  4  & -        & $^1 S         $ & 137 & 29 & 339 \\
B      &  5  & -        & $^2 P         $ & 521 & 43 & 368 \\
C      &  6  & -        & $^3 P         $ & 550 & 52 & 377 \\
N      &  7  & -        & $^4 S         $ & 458 & 25 & 350 \\
O      &  8  & -        & $^3 P         $ & 472 & 41 & 366 \\
F      &  9  & -        & $^2 P         $ & 553 & 29 & 354 \\
Ne     & 10  & -        & $^1 S         $ & 499 & 14 & 324 \\ \hline
Li$_2$ &  6  & 5.051    & $^1 \Sigma^+_g$ & 120 & 34 & 275 \\
Be$_2$ &  8  & 4.65     & $^1 \Sigma^+_g$ &  39 & 11 & 252 \\
B$_2$  & 10  & 3.005    & $^3 \Sigma^-_g$ &  61 & 16 & 257 \\
C$_2$  & 12  & 2.3481   & $^1 \Sigma^+_g$ &  69 & 17 & 258 \\
N$_2$  & 14  & 2.075    & $^1 \Sigma^+_g$ &  35 & 10 & 251 \\
O$_2$  & 16  & 2.283    & $^3 \Sigma^-_g$ &  12 &  8 & 249 \\
F$_2$  & 18  & 2.668    & $^1 \Sigma^+_g$ &   7 &  6 & 247 \\
Ne$_2$ & 20  & 5.84     & $^1 \Sigma^+_g$ &   6 &  5 & 246 \\ \hline
LiH    &  4  & 3.0139   & $^1 \Sigma^+  $ &  31 & 15 & 452 \\
BeH    &  5  & 2.5372   & $^2 \Sigma^+  $ &  30 & 18 & 455 \\
CH     &  7  & 2.1163   & $^2 \Pi       $ &  25 & 13 & 450 \\
NH     &  8  & 1.9581   & $^3 \Sigma^-  $ &  21 &  9 & 446 \\
OH     &  9  & 1.8324   & $^2 \Pi       $ &  25 & 13 & 450 \\
FH     & 10  & 1.7328   & $^1 \Sigma^+  $ &  61 & 13 & 450 \\ \hline
LiF    & 12  & 2.9553   & $^1 \Sigma^+  $ &  62 & 14 & 451 \\
CN     & 13  & 2.2144   & $^2 \Sigma^+  $ &  35 &  8 & 445 \\
CO     & 14  & 2.1321   & $^1 \Sigma^+  $ &  37 &  9 & 446 \\
NO     & 16  & 2.1746   & $^2 \Pi       $ &  53 & 14 & 451 \\ \hline \hline
\end{tabular}
\caption{ \label{tab:2}
Summary of parameters for the isolated atoms and diatomic molecules.
Experimental bond lengths used are as Ref.~\cite{toulouse08} for the homogeneous diatomic molecules, and from Ref.~\cite{ruette05} for the rest.
}
\end{table}

\begin{table}[t]
\begin{tabular}{lr@{.}lr@{.}lr@{.}l}    \hline \hline
        & \multicolumn{2}{c}{$E_{vmc}(a.u.)$}  & \multicolumn{2}{c}{$E_{exp}(a.u.)$} & \multicolumn{2}{c}{$E_{corr}(\%)$}  \\ \hline
Li      &       -7&478052(2)   &       -7&47806032  &    99&981(5)  \\
Be      &      -14&667243(3)   &      -14&66736     &    99&876(4) \\
B       &      -24&65329(1)    &      -24&65391     &    99&503(8) \\
C       &      -37&84361(2)    &      -37&8450      &    99&11(1) \\
N       &      -54&58641(4)    &      -54&5892      &    98&52(2) \\
O       &      -75&06058(5)    &      -75&0673      &    97&39(2) \\
F       &      -99&72623(8)    &      -99&7339      &    97&64(3) \\
Ne      &     -128&9299(1)     &     -128&9376      &    98&02(3) \\ \hline
Li$_2$  &      -14&98398(4)    &      -14&9951      &    91&00(3) \\
Be$_2$  &      -29&31885(7)    &      -29&3380      &    90&60(4) \\
B$_2$   &      -49&3837(1)     &      -49&415       &    90&34(4) \\
C$_2$   &      -75&8893(2)     &      -75&9265      &    92&83(4) \\
N$_2$   &     -109&4981(3)     &     -109&5421      &    91&99(6) \\
O$_2$   &     -150&2843(4)     &     -150&3267      &    93&57(6) \\
F$_2$   &     -199&4819(4)     &     -199&5303      &    93&60(5) \\
Ne$_2$  &     -257&8517(5)     &     -257&8753      &    96&99(6) \\  \hline
LiH     &       -8&06908(2)    &       -8&0704      &    98&41(2) \\
BeH     &      -15&24407(2)    &      -15&2468      &    97&08(3) \\
CH      &      -38&46842(5)    &      -38&4788      &    94&78(3) \\
NH      &      -55&20863(8)    &      -55&2227      &    94&24(3) \\
OH      &      -75&7222(1)     &      -75&7371      &    95&29(4) \\
FH      &     -100&4446(2)     &     -100&4592      &    96&24(4) \\ \hline
LiF     &     -107&4177(2)     &     -107&4344      &    96&22(4) \\
CN      &      -92&6755(2)     &      -92&7250      &    90&10(5) \\
CO      &     -113&2845(3)     &     -113&3261      &    92&23(6) \\
NO      &     -129&8457(3)     &     -129&9047      &    90&27(5) \\ \hline \hline
\end{tabular}
\caption{ \label{tab:3}
Total electronic energies for isolated atoms and diatomic molecules.
The table presents VMC estimates obtained as described in this paper ($E_{VMC}$).
Approximate exact total energies for isolated atoms are from Ref.~\cite{puchalski06} for Li and Ref.~\cite{chakravorty93} otherwise.
Approximate exact total energies are as Ref.~\cite{toulouse08} for Be$_2$,B$_2$,C$_2$,Ne$_2$, and from Ref.~\cite{oneill05} otherwise.
}
\end{table}

Table~\ref{tab:3} shows the resulting total energy estimates.
For comparison of these results with other methods we focus on the variation in the fraction of total correlation energy recovered for each system, taking
RHF energies from numerical calculations using \textsc{ATSP2K} and \textsc{2dhf}, and approximate exact total energies from the sources specified in the table.

The lowest variational total energy estimates for first row isolated atom in the literature appears to be those of Brown \emph{et al.} \cite{brown07}.
They employ standard sampling, linear optimisation, and trial wavefunction of the same form as Eq.~(\ref{eq:wf}) but with less variational freedom, mostly due to a smaller multi-determinant expansion.
Our VMC results consistently fall between the VMC and DMC estimates of this paper, are consistently closer to DMC than VMC estimates of this paper, and provide a small but significant recovery of between 0.1\% and 1.5\% more correlation energy at VMC level.

Toulouse and Umrigar\cite{toulouse08} also provide isolated atom energies using both VMC and DMC, based on linear optimisation of trial wavefunction using standard sampling.
Their trial wavefunctions were composed of RHF or MCSCF orbitals represented using a Slater basis set, with orbital relaxation, a Jastrow factor, and 1 or 2 CSFs.
Our VMC energies consistently improve on the estimates in this paper, providing between 0.3\% and 12.5\% more correlation energy than the VMC estimates, and between 0.0\% and 5.6\% more correlation than DMC estimates.
The improvement is entirely consistent with the contribution to correlation provided by backflow and multi-determinant expansions discussed in Ref.~\cite{brown07}.

Toulouse and Umrigar\cite{toulouse08} also provide VMC and DMC total energies for the first row homonuclear diatomic molecules, with trial wavefunctions constructed in the same manner as the isolated atoms but allowing larger multi-determinant expansions for some of the molecules.
The difference between results is not as consistent as for the isolated atoms.
Our VMC results recover between 6.7\% less and 5.4\% more correlation energy, with a consistent inprovement as electron number is increased and the crossover occuring between B$_2$ and C$_2$.
The same trend occurs when comparing with DMC results, with our VMC energies higher than the DMC results of Ref.~\cite{toulouse08} for all molecules except Ne$_2$.
This is consistent with the isolated atom behaviour and the observation that our multi-determinant expansion is inferior in that it is not obtained from a self-consistent calculation.
This reduction in correlation energy is apparent for the small systems only, with the contribution to correlation energy provided by backflow becoming dominant as electron number increases.

Curtiss \emph{et al.}\cite{curtiss07} provide total energies for a large set of molecules arrived at using the Gaussian-4 (G4) theory, a state-of-the-art perturbation method.
These provide a useful context for the VMC results presented here, provided we bear in mind that the comparison is less clear cut since G4 errors may be positive or negative due to it being a non-variational method.
Of the 26 VMC results in this paper G4 total energies are available for 23.
Of these, our VMC total energies are below (above) those of G4 for 17 (6) of the systems, with the VMC recovering between 25.8\% more and 4\% less correlation energy than the G4 results.
The isolated atom energies are consistently lower for VMC, but beyond this very little consistent behaviour is evident beyond the general observation that VMC energies are lower more often than not, and where the VMC energy is greater than G4 the difference is small.

Finally, we compare the correlation recovered by VMC between the systems considered.
Perhaps the most striking feature is the superior performance for isolated atoms when compared with the diatomic molecules.
This can partly be ascribed to the superior determinants used for the isolated atoms when compared with diatomic molecules (the latter are made up of fewer determinant and are not self-consistent at the multi-determinant level), but it seems unlikely that this is enough when we consider the wide variation in the amount of correlation recovered.
A rough observation is that correlation is most effectively described for those systems where the electronic behaviour around a nucleus is most similar to an isolated atom or ion, for example for the ionic hydrides such as LiH, or the very weakly bonded molecules such as Ne$_2$.
This suggests the possibility that `missing correlation' may well be due to Jastrow and backflow being limited to radial functions, and that it may be useful to replace these with functions that allow more geometric freedom.
This must inevitably involve introducing more parameters, but the combination of efficient sampling, linear optimisation, and parameter averaging employed here appears to be able to handle large parameter sets reliably and efficiently.

\section{Conclusions}
\label{sec:conc}

Fermionic Variational quantum Monte Carlo is almost always implemented by drawing sample values of particle co-ordinates from the distribution $P_{s}=\lambda \psi^2$, where $\psi$ is the trial wavefunction whose expectation values are sought - the `standard sampling' of this paper.
We have shown that this is an \emph{ad hoc} choice, that the only special property of this choice is the simplicity of the resulting expressions for estimates, and that it has some undesirable properties.

An analysis of VMC using a general sampling distribution is described.
We provide the conditions that must be satisfied for the resulting VMC estimates to be Normally distributed, conditions that are easily violated by sample distributions that unnecessary possesses a nodal surface.
For the case where these conditions are satisfied, general estimates are derived for the parameters of the underlying Normal distribution, so providing estimates, confidence limits, and error bars.
When the conditions are not satisfied, the underlying distributions are shown to be of a known but undesirable form that are heavy tailed and that are not characterised by a mean and variance.
Obtaining estimates of the parameters of such distributions is considerably more difficult than the Normal case, and even if such estimates were available the use of error-bars is less informative than the Normal case.

This analysis is extended to VMC optimisation in a less rigorous manner to provide similar conditions that must be satisfied in order that the random fluctuation in optimised parameters is Normal,
and that the error in a total energy due to the randomness of these parameters possess a mean and variance.
Applying these conditions to the special case of standard sampling informs us that although VMC total energy estimates for fixed parameters are Normally distributed, almost all of the quantities that appear in optimisation are not.

Two new sampling strategies are presented, both of which ensure a Normal random error wherever possible.
Optimum sampling is obtained by deriving the sample distribution for which the Normal random error is a small as possible for a given sample size.
This naturally provides a lower limit to the random error possible for a given system, and an accurate approximation to this distribution is derived.

Efficient sampling is designed by attempting to avoid a large part of the the computational cost of providing samples using the Metropolis algorithm.
The resulting sample distribution function is computationally cheap to evaluate, has the correct properties for estimates and optimisation functions to be Normally distributed, and does not perform significantly worse than standard or optimum sampling for a given samples size.

Of the two, efficient sampling provides the most useful improvement over standard sampling by increasing the number of samples possible for a fixed computational resource by an order of magnitude and significantly decreasing the available random error.
In addition, the analysis of optimum and efficient sampling suggests improved statistical properties for optimisation when compared to standard sampling.
However, no convincing numerical evidence that this is the case for real systems was obtained, mainly due to the greater computational cost of standard sampling, the small random error in optimisation compared to estimation, and that any rigorous test would require a statistically significant number of optimisation calculations to be compared.
Perhaps the most useful results provided by this analysis is a theoretical justification for averaging sets of optimised parameters in order to reduce the optimisation random error.

Variational quantum Monte Carlo using efficient sampling and optimisation was applied to evaluate the electronic total energy for 26 isolated atoms and molecules, using trial wavefunction with a great deal of variational freedom and a fixed and modest computational budget divided equally amongst calculations.
At VMC level more than 97\% of correlation energy was recovered for first row isolated atoms, and more than 90\% for a small set of diatomic molecules, results that significantly improve on many previously published VMC results and approaching DMC for some systems.

As systems increase in size the efficient sampling strategy is likely to become less advantageous, but this has not occurred for the systems considered so far.
When this does occur it should be possible to employ further sampling distributions that perform better for these systems such as optimum sampling, or the distribution used by Attaccalite and Sorella\cite{sorella08} for a dense extended system.

\begin{acknowledgements}
The authors thank R.J. Needs of the University of Cambridge for many useful discussions.
Financial support was provided by Special Coordination Funds for Promoting Science and Technology, 
``Promotion of environmental improvement to enhance Young Researchers' independence, and make use of their abilities'' (``Development of Personnel System for Young Researchers in Nanotechnology and Materials Science'' (Japan Advanced Institute of Science and Technology)) 
and by a Grant--in--Aid for Scientific Research in Priority Areas ``Development of New Quantum Simulators and Quantum Design (No. 17064016)'' (Japanese Ministry of Education,Culture, Sports, Science, and Technology ; KAKENHI-MEXT).
\end{acknowledgements}

\appendix
\section{Power law tails from singularities at the nodal surface}
\label{sec:appa}

Throughout this paper it is repeatedly stated that in order to characterise the form of the distribution of an estimate all that is required is a knowledge of the behaviour of a sample and its distribution in the region of the nodal surface of the trial wavefunction.
This appendix describes how this is done, and is a summary and generalisation of the method applied to the standard case in a previous publication\cite{trail08a}.

A MC estimate is constructed by drawing $r$ samples from a distribution in $3N$ dimensional co-ordinate space, $P(\mathbf{R})$, evaluating a functional at each of these samples, $x_L(\mathbf{R}_i)$, and then taking a sample mean, that is the integral
\begin{equation}
X=\int P(\mathbf{R}) x_L(\mathbf{R}) d^{3N}\mathbf{R}
\end{equation}
is estimated as a sample value of the random variable $\mathsf{X}$ given by 
\begin{equation}
\mathsf{X} = \frac{1}{r} \sum{ x_L( \text{\sffamily\bfseries R} ) }.
\end{equation}
It is necessary for the central limit theorem to be valid in order to know the distribution of $\mathsf{X}$ and to characterise the random error in terms of a confidence intervals for a particular value of the estimate.
A valid CLT requires the first and second moments of $\mathsf{x}=x_L( \text{\sffamily\bfseries R} )$ to exist, where $\mathsf{x}$ is a random variable whose distribution is critically dependant on the presence of singularities in the function $x_L( \mathbf{R} )$.

The distribution of $\mathsf{x}$ can be obtained from that of $\text{\sffamily\bfseries R}$ using the standard formulae
\begin{equation}
P(x)=\int_{x=x_L(\mathbf{R})} \frac{ P(\mathbf{R}) }{ |\nabla_{\mathbf{R}} x_L | } d^{3N-1}\mathbf{R},
\label{eq:b1}
\end{equation}
where the integral is taken over all surfaces of constant $x_L=x$.
Here we consider the case where $x_L$ and $P(\mathbf{R})$ possess singularities and/or zeroes on a $3N-1$ dimensional hyper-surface, the particular case that repeatedly occurs within QMC due to the existence of the nodal surface in fermionic wavefunctions.

Introducing a general curvilinear co-ordinate system in terms of surfaces of constant $x_L$ provides a new co-ordinate system $(s,\mathbf{T})$, with $s$ a scalar set to zero on the $3N-1$ dimensional hyper-surface of interest and $\mathbf{T}$ an implicit $3N-1$ vector contained in the hyperplane tangential to the constant $x_L$ surface at $\mathbf{R}$.
Expanding the distribution about the hyper-surface of interest (usually the nodal surface of some trial wavefunction) in this new co-ordinate system gives
\begin{equation}
P(s,\mathbf{T}) = s^m \left[a_0(\mathbf{T}) + a_1(\mathbf{T})s + \ldots \right]
\end{equation}
and analogously $x_L$ becomes
\begin{equation}
x_L(s,\mathbf{T}) = \frac{1}{s^n} \left[b_0 + b_1s + \ldots \right]
\end{equation}
with $(m,n)$ characterising the zeros and singularities of $P$ and $x_L$.

In terms of $(s,\mathbf{T})$ the distribution of $x$ is given by
\begin{equation}
P(x)=\int  P(s,\mathbf{T}) \left| \frac{d s}{d x_L} \right| d^{3N-1}\mathbf{T},
\end{equation}
so in the limit $s\rightarrow0$
\begin{equation}
P(x) \propto \frac{s^m}{1/|s^{n+1}|}
\end{equation}
and in the limit $x\rightarrow \pm \infty$
\begin{equation}
P(x) \asymp \left( \frac{1}{|x|} \right)^{\frac{m+n+1}{n}}.
\end{equation}
The existence of positive and/or negative tails, and their possible equality, is entirely decided by the continuity of $P(\mathbf{R})$ at the hyper-surface of interest, and the symmetry of the singularity in $x_L$.
An example of this analysis arises for the total energy estimate via standard sampling, for which $(m,n)=(2,1)$, hence $P$ possesses asymptotic tails that decay as $x^{-4}$, and the CLT is valid for the estimate since the second moment exists.

For the generalised sampling a Normally distributed estimate requires a valid bivariate CLT, and this is decided by the existence of the covariance matrix.
The random estimate is 
\begin{equation}
\mathsf{Z}=\frac{\mathsf{Y}}{\mathsf{X}} =
\frac{ \sum \mathsf{y}_i }{ \sum \mathsf{x}_i }
\end{equation}
with each $\{ \mathsf{y}_i \}$ and $\{ \mathsf{x}_i \}$ independent and identically distributed, with variables in the two sets independent for $i \neq j$, co-distributed as $P(y,x)$ for $i=j$.
The validity of the bivariate CLT requires the covariance matrix of $P(y,x)$ to exist, and since this matrix is positive definite we only require all diagonal matrix elements to exist - the variance of both $\mathsf{y}_i$ and $\mathsf{x}_i$.
Consequently, we need only consider the distribution of each random variable independently of the other, given by Eq.~(\ref{eq:b1}) for $\mathsf{x}_i$ and the equivalent expression for $\mathsf{y}_i$, and the univariate analysis given above can be applied.
Note that $\mathsf{y}_i$ and $\mathsf{x}_i$ are parametrically related to each other (via the underlying random spatial variable) so $P(y,x)$ is non-zero only on a line in two dimensional space, but this does not prevent the existence of the covariance matrix or the continuity of $P(Y,X)$ as long as more than $1$ sample appears in each sum.

An example arises for optimisation with standard sampling, for which $(m,n)=(2,2)$ for both of $(\mathsf{y}_i,\mathsf{x}_i)$ as soon as the nodal surface moves away from its starting position.
So, $P(y,x)$ possesses positive asymptotic tails that decay as $y^{-\frac{5}{2}}$ and $x^{-\frac{5}{2}}$ for $(\mathsf{y}_i,\mathsf{x}_i)$, neither of the variances are defined, the covariance matrix is not defined, and the estimate $\mathsf{Z}$ is not Normal in the large $r$ limit.
Generally, for $P(y,x)$ decaying slower than either $|x|^{-3}$ or $|y|^{-3}$ in the asymptotic limit a bivariate Stable distribution results, and although a generalisation of Feller's theorem could be constructed it cannot provide a Normally distributed estimate for the quotient, or estimates of confidence limits.


\begin{thebibliography}{}
\bibitem{foot1}        Finite $\sigma_{opt}$ for an infinite density matrix may occur by a cancellation of non-integrable singularities in the sum of integrals in Eq.~(\ref{eq:3}), $\left[c_{22} - 2 \left( \frac{\mu_2}{\mu_1} \right)c_{12} + \left( \frac{\mu_2}{\mu_1} \right)^2c_{22}\right]$.
\bibitem{foot2}         For standard sampling we replace $P_g$ with $\psi^2$, and may replace $m$ with $(m-1)$.
\bibitem{foot3}         These are biased estimates. The bias is small and equal to $\widehat{\sigma_E^2}/r$ defined by Eq.~(\ref{eq:5}), provided the bivariate CLT is valid. We do not use unbiased estimates to define the minimised quantity since they do not change the discussion in this section, but do complicate the equations for general sampling.
\bibitem{foot4}         This is not entirely obvious since Eq.~(\ref{eq:10}) includes a total energy estimate, hence the averaged variables are neither independent or identically distributed.  However, the bivariate CLT remains valid due to the contribution of correlation between samples at different times decaying relative to the contribution of samples at the equal times, as sample size increases.  A derivation is analogous to showing that variance estimates become Normal.
\bibitem{foot5}         The computational budget for each system was 24h of computational time for monitored optimisation (12h for optimisation, 12h for sample generation) followed by 24h for the final accurate estimate, with calculations performed on a modest desktop with 2 quad-core processors.
\end{thebibliography}
\end{document}